# Distributed Time-Frequency Division Multiple Access Protocol For Wireless Sensor Networks

Dujdow Buranapanichkit and Yiannis Andreopoulos[*], *Member, IEEE*

*Abstract*—It is well known that biology-inspired self-maintaining algorithms in wireless sensor nodes achieve near optimum time division multiple access (TDMA) characteristics in a *decentralized* manner and with *very low complexity*. We extend such distributed TDMA approaches to multiple channels (frequencies). This is achieved by extending the concept of collaborative reactive listening in order to balance the number of nodes in all available channels. We prove the stability of the new protocol and estimate the delay until the balanced system state is reached. Our approach is benchmarked against single-channel distributed TDMA and channel hopping approaches using TinyOS imote2 wireless sensors.

## I. INTRODUCTION

DISTRIBUTED coordination for collision-free packet transmissions in networks of wireless sensor nodes is a long-standing research problem [1]-[4]. A critical aspect of this problem is *distributed synchronization*, which enables wireless sensor nodes to easily and efficiently build a time division multiple access (TDMA) mechanism that minimizes collisions. Recently, schemes based on pulse-coupled oscillators (PCOs) [3][4] gained attention for the dual (equivalent) problem, i.e. *desynchronization*. The basic premise of these methods is *reactive listening for TDMA* within a decentralized medium access control (MAC) protocol. In comparison to centralized (de)synchronization, such schemes are robust to clock drift and transmission delay jitter and, importantly, do not require the presence of a coordinating node.

Complementary to desynchronization for distributed TDMA, multi-channel MAC protocols aim for load balancing via *frequency division multiple access* [5]-[8], or TDMA combined with pseudo-random channel hopping, e.g. as proposed for the upcoming IEEE 802.15.4e standard [9]. Their key principles are: *(i)* collection of traffic statistics or TDMA coordination by a central station; *(ii)* centralized channel assignment (or hopping) for interference reduction [2]. A protocol forming an exception to these principles is EM-MAC [8], which allows for distributed multichannel coordination with predictive wake-up scheduling.

In this paper we propose distributed MAC-layer *time-frequency division multiple access* (TFDMA) for wireless sensor networks (WSNs) based on reactive listening of message broadcasts. Unlike previous TFDMA schemes [5]-[7] that are centralized or highly-complex for real-world sensor devices (due to complex heuristics or NP-time algorithms), our approach forms a low-complex *decentralized* scheme based on reactive listening. Unlike channel hopping approaches like EM-MAC [8] or schemes based on the IEEE 802.15.4e MAC [9][10], we avoid continuous channel switching and achieve a significantly smaller network setup delay and higher bandwidth efficiency. This makes our proposal suitable for WSN-based monitoring applications requiring rapid network setup and high data throughput once an alert is triggered. Beyond the proposed TFDMA, this paper's theoretical contributions are: *(i)* we prove that distributed TFDMA converges to steady state (SS) under appropriate parameter setting; *(ii)* we derive, and validate, the expected delay for convergence to SS.

## II. SUMMARY OF DESYNCHRONIZATION FOR DISTRIBUTED TDMA

Consider a network of fully-connected wireless sensors[1] (or "nodes"). A message broadcasted from a node is received from all other nodes. All nodes broadcast one short beacon message within periodic intervals of *T*s. The mechanism described here follows the DESYNC protocol [4]. A PCO-based variation with limited listening time per period was proposed recently [3]; it can be used in our work with minor adjustments.

Each node $n_{\text{curr}}$ (out of $W_{\text{tot}}$ nodes) picks a particular time instant $t_{\text{curr}}$ to broadcast its beacon based on the previous and next beacon broadcasts (stemming from nodes $n_{\text{previous}}$ and $n_{\text{next}}$). The determination of this time instant is performed immediately after $n_{\text{curr}}$ receives the beacon of $n_{\text{next}}$. Hence, $n_{\text{curr}}$ listens to all other nodes' beacon broadcasts and, during the *k*th iteration (period), schedules its next beacon time according to the reactive listening primitive [4]:

$$t_{\text{curr}}^{(k+1)} = T + (1-\alpha)t_{\text{curr}}^{(k)} + \alpha \frac{t_{\text{previous}}^{(k)} + t_{\text{next}}^{(k)}}{2}, \quad (1)$$

where $T$ is the desired TDMA period (in s) and $\alpha \in (0,1)$ is a parameter that scales how far $n_{\text{curr}}$ moves from its current beacon time (at $t_{\text{curr}}^{(k)}$) toward the desired midpoint [3][4].

Previous work [3][4] showed that the reactive listening primitive of (1) leads to near-optimal TDMA behavior in SS,

[*]Corresponding author. The authors are with the Electronic and Electrical Engineering Department, University College London, Roberts Building, Torrington Place, London, WC1E 7JE, Tel. +44 20 7679 7303 (both authors), Fax. +44 20 7388 9325 (both authors), Email: d.buranapanichkit@ee.ucl.ac.uk (D. Buranapanichkit), iandreop@ee.ucl.ac.uk (Y. Andreopoulos).

This work has been partially supported by the EC under contract FP7-2007-IST-2-224053 (CONET project) and by a PhD scholarship from the Government of Thailand. This version corresponds to the accepted paper that may differ from the published paper due to size limitations of the journal.

---

[1] Notations: $n_{\text{curr}}$ is the sensor node under consideration; $t_{\text{curr}}$ is the time instant $n_{\text{curr}}$'s beacon is broadcasted (we ignore propagation and system delays); $W_{\text{tot}}$ is the total number of nodes and $W_c$ is the number of nodes operating in channel $c$ (Ch{$c$}); $t_{\text{curr}}^{(k)}$ is quantity $t_{\text{curr}}$ computed after $k$ iterations; $\overline{W}$ is the expected value of $W$; $0.\overline{9} = 0.999...$; $\lfloor a \rfloor$, $\lceil a \rceil$ & $[\![a]\!]$ indicate floor, ceiling & rounding.

i.e. after $k_{ss}$ periods, where all beacon messages are periodic with:

$$\left|t_{\text{curr}}^{(k_{ss}+1)} - t_{\text{curr}}^{(k_{ss})} - T\right| < q_{ss}T, \quad (2)$$

with $q_{ss}$ a preset threshold, e.g. $q_{ss} = 0.02$. In SS, each node transmits data packets for $T/W_{\text{tot}}$s immediately following its beacon-message broadcast. If a node joins or leaves the network, thereby leading to $W_{\text{tot}}' \neq W_{\text{tot}}$ beacon-message broadcasts, the remaining nodes reconfigure their beacon-message broadcasts to converge to a new TDMA state and then continue data transmission once (2) is satisfied. Once TDMA behavior is achieved, the only overhead stems from the beacon-message broadcasts, which include the node's number. Experiments can be performed to establish the expected value of $k_{ss}$ [3][4]. Beyond fully-connected WSNs, DESYNC has been extended to multi-hop topologies [11] and convergence to SS has been also proven for this case.

## III. PROPOSED MULTI-CHANNEL EXTENSION OF DISTRIBUTED TDMA DESYNCHRONIZATION

Standards suitable for WSNs, such as the IEEE 802.15.4 MAC, allow for half-duplex communications over a selection of channels at 2.4GHz with minimal cross-channel interference. This hints that, should TDMA desynchronization be extended to $C$ channels ($C > 1$), increased throughput per node will be observed since $[\![W_{\text{tot}}/C \pm 0.4\overline{9}]\!]$ nodes will operate in each channel. The highest throughput can be achieved when the number of nodes is balanced in all channels [5]. For example, for $C = 2$, the aim would be to "spontaneously" separate $W_{\text{tot}} = 8$ nodes into two distinct sets: $W_1 = W_2 = 4$, i.e. 4 nodes in each channel. This uses the allocated spectrum of IEEE802.15.4 twice as efficiently in comparison to PCO-based TDMA [3][4]. However, channel switching must be designed judiciously, as frequent channel switching causes loss of (de)synchronization due to variable hardware and operating system latencies and additional effort (and energy consumption) is required to recover it [8].

### A. Proposed Protocol

By utilizing reactive listening, TFDMA only allows for channel switching if less nodes are detected in the new channel. The detailed operation is described here.

**Switching:** In the beginning, each wireless sensor picks a channel Ch$\{c\}$ ($1 \leq c \leq C$) randomly and applies DESYNC [4]. After broadcasting its beacon message, each node can switch to the previous or next channel, i.e. from Ch$\{c\}$ to Ch$\{c + s_c\}$ ($1 \leq c \leq C$, with $s_c \in \{\pm 1, ..., \pm \lfloor C/2 \rfloor\}$ and cyclic extension: Ch$\{C + |s_c|\} \equiv$ Ch$\{|s_c|\}$, Ch$\{1 - |s_c|\} \equiv$ Ch$\{C + 1 - |s_c|\}$), by broadcasting a "switch" message in Ch$\{c\}$. This message contains the node number and alerts all other nodes listening and transmitting in Ch$\{c\}$ that this node will attempt to switch to a different channel. Once receiving one switch message, all other nodes in Ch$\{c\}$ disable the desynchronization process and, instead of assigning their next beacon-message broadcast based on (1), they simply repeat it after $T$s for the next two periods. This is termed "switch" mode.

**Reactive listening:** The node switching to Ch$\{c + s_c\}$ listens to the beacon messages of Ch$\{c + s_c\}$ for one period[2] and determines if $W_{c+s_c} \leq W_c - 2$. If so, it joins the new channel and distributed TDMA is achieved in Ch$\{c\}$ and Ch$\{c + s_c\}$ via DESYNC. Otherwise it returns to Ch$\{c\}$, broadcasts a "return" message, and rejoins desynchronization and data transmission in Ch$\{c\}$. Nodes in Ch$\{c\}$ exit the switch mode and continue their regular desynchronization operation when a return message is received, or after two periods.

Assuming $s_c^{(k)} > 0$ for the $k$th switch mode of Ch$\{c\}$, if a return message is received, all nodes in Ch$\{c\}$ set $s_c^{(k+1)} = -s_c^{(k)}$, i.e., when unsuccessful, the switching direction changes; furthermore $s_c$ gradually increases up to $\pm \lfloor C/2 \rfloor$ to cover all channels. An update occurring simultaneously between channels: $c \to c + s_c^{(k)}$ and $\acute{c} \to c$ ($1 \leq \acute{c} \leq C$ & $\acute{c} \neq c$) is expressed stochastically for Ch$\{c\}$ by:

$$\overline{W}_c^{(k+1)} = \overline{W}_c^{(k)} - \min\left\{u\left[\overline{W}_c^{(k)} - 2 - \overline{W}_{c+s_c^{(k)}}^{(k)}\right]p_{\text{sw},c}^{(k)}\overline{W}_c^{(k)}, 1\right\} \\ + \min\left\{u\left[\overline{W}_{\acute{c}}^{(k)} - 2 - \overline{W}_c^{(k)}\right]p_{\text{sw},\acute{c}}^{(k)}\overline{W}_{\acute{c}}^{(k)}, 1\right\}, \quad (3)$$

with $u[\cdot]$ the unit-step function, used to identify whether switching can occur between channels $c \to c + s_c^{(k)}$ and $\acute{c} \to c$, and $p_{\text{sw},c}^{(k)}, p_{\text{sw},\acute{c}}^{(k)}$ the switching probabilities of a node in Ch$\{c\}$ and Ch$\{\acute{c}\}$.

**Stability and convergence mechanism:** Since each node decides and sends its switch message immediately after its beacon message, once one such message is heard in one period, the remaining nodes in that channel cannot switch in this period. The switch mode allows for undisturbed operation while nodes find out if the previous or next channel has less nodes: *(i)* if a node returns, it can quickly regain its previous slot with minimal disturbance; *(ii)* via the switch mode, the reactive listening primitive of (3) is used for adjustment of the number of nodes per channel. Once the switch mode is exited for the $k$th time in Ch$\{c\}$, each node modifies its switching probability by:

$$p_{\text{sw},c}^{(k)} = \min\left\{\beta^v \times p_{\text{sw},c}^{(k-1)}, 1\right\}, \quad (4)$$

where: $v = 1$ if no return message is received, $v = -1$ otherwise, and $\beta > 1$. Initially, each node of channel Ch$\{c\}$ will attempt to switch with probability $p_{\text{sw},c}^{(0)}$ (which is preset); $\beta$ controls the "back-off" from switching (also preset), and $v$ changes according to the result of the last switch attempt.

Notice that, once $[\![W_{\text{tot}}/C \pm 0.4\overline{9}]\!]$ nodes exist in all channels, further switching attempts will cause the nodes to return to their original channel, thus leading to $\forall c: p_{\text{sw},c}^{(\text{ss})} \to 0$ from (4). Thus, even in steady state we enforce infrequent channel switching attempts to periodically discover and compensate for potential imbalances created by nodes departing unexpectedly (e.g. if nodes malfunction): we impose that a node in each channel will attempt to switch after $Z$ periods of switching inactivity.

Both the periodic beacon message broadcasts and the reactive listening principle are of critical importance for (3) and for the proposed TFDMA operation as they ensure switching nodes can detect the number of nodes in the new channel (and whether the new channel is in fact in switch mode). We provide a TinyOS nesC implementation of the proposed distributed TFDMA online [12].

---

[2] Each beacon message includes the total number of nodes heard in Ch$\{c\}$, as well as a flag indicating whether the channel is in switch mode (i.e. whether a node has left to listen to Ch$\{c + s_c\}$). Thus, each node finds $W_c$ (and whether switch mode is on) even if only a single beacon message is heard in Ch$\{c\}$.

*B. Theoretical Analysis*

*Proposition 1 (Convergence to SS):* An arbitrary distribution of $W_{\text{tot}}$ nodes in $C$ channels ($W_{\text{tot}} \geq 2C$) will be driven to balanced state of $[\![W_{\text{tot}}/C \pm 0.4\overline{9}]\!]$ nodes per channel under TFDMA with $0 < \alpha < 1$.

*Proof:* Single-channel TDMA desynchronization has already been shown to converge for $0 < \alpha < 1$ [3][4]. Thus, it suffices to show that the proposed channel switching mechanism leads to balanced number of nodes per channel.

For every channel $c$ ($1 \leq c \leq C$), when $s_c^{(k)} = 1$ the transition system formed by (3) for all $C$ channels is written in matrix form as:

$$\overline{\mathbf{w}}^{(k+1)} = \mathbf{G}^{(k)} \overline{\mathbf{w}}^{(k)} \quad (5)$$

with

$$\overline{\mathbf{w}}^{(k+1)} = \begin{bmatrix} \overline{W}_1^{(k+1)} & \overline{W}_2^{(k+1)} & \cdots & \overline{W}_{C-1}^{(k+1)} & \overline{W}_C^{(k+1)} \end{bmatrix}^T \quad (6)$$

$$\overline{\mathbf{w}}^{(k)} = \begin{bmatrix} \overline{W}_1^{(k)} & \overline{W}_2^{(k)} & \cdots & \overline{W}_{C-1}^{(k)} & \overline{W}_C^{(k)} \end{bmatrix}^T \quad (7)$$

$$\mathbf{G}^{(k)} = \begin{bmatrix} 1-g_1 & 0 & \cdots & 0 & g_C \\ g_1 & 1-g_2 & \cdots & 0 & 0 \\ \vdots & \vdots & \ddots & \vdots & \vdots \\ 0 & 0 & \cdots & 1-g_{C-1} & 0 \\ 0 & 0 & \cdots & g_{C-1} & 1-g_C \end{bmatrix} \quad (8)$$

and $\forall c: g_c = u[\overline{W}_c^{(k)} - \overline{W}_{c+1}^{(k)} - 2]p_{\text{sw},c}^{(k)}$ with the constraint of $\forall c: g_c \overline{W}_c^{(k)} \leq 1$ due to the $\min\{\cdot\}$ operators of (3).

For the general case of $s_c^{(k)} \neq 1$, factors $g_c$ of $\mathbf{G}$ are positioned in column $c$ and row $c + s_c^{(k)}$, with cyclic extension at the borders (i.e. when $c + s_c^{(k)} > C$ or $c + s_c^{(k)} < 1$). The stochastic transition matrix $\mathbf{G}$ of (5) under any $s_c^{(k)}$ is a left-stochastic matrix with: its columns maximally summing to unity, all its entries being non-negative and each entry is smaller or equal to unity. As such, via the Perron–Frobenius theorem [13], we find that the maximum magnitude of all eigenvalues of $\mathbf{G}$ is unity, i.e. all eigenvalues of any instantiation of $\mathbf{G}$ are within (or on) the unit circle. Hence, under iterations with stochastic matrices $\mathbf{G}$, the system of (5) will converge to a steady state or to a limit cycle. Limit cycles, i.e. oscillations between unbalanced numbers of nodes per channel, are avoided since, under the reactive listening of Section III.A [expressed stochastically by (3)], nodes switch only if they join a channel with less nodes. The inclusion of the total number of nodes (and switch mode status) of each channel within *each* beacon message (see footnote 2) ensures that no erroneous node switching can occur during convergence to SS even under the occasional loss of a switch or beacon message. Hence, the system of (5) will converge to a steady state. All vectors:

$$\mathbf{w}^{(\text{SS})} = [[\![W_{\text{tot}}/C \pm 0.4\overline{9}]\!] \quad \cdots \quad [\![W_{\text{tot}}/C \pm 0.4\overline{9}]\!]]^T \quad (9)$$

comprise the eigenvectors (fixed points) of the system of (5) and lead to $\mathbf{G} = \mathbf{I}$ (i.e. they all correspond to unity eigenvalues). This is because all $\mathbf{w}^{(\text{SS})}$ of (9) lead to:

$$\forall x, y \in \{1, \dots, C\}: \max\{|\overline{W}_x^{(k)} - \overline{W}_y^{(k)}|\} = 1$$

$$\Rightarrow \forall x, y: u[\overline{W}_x^{(k)} - 2 - \overline{W}_y^{(k)}] = u[\overline{W}_y^{(k)} - 2 - \overline{W}_x^{(k)}] = 0$$

$$\Rightarrow \forall c: g_c = 0.$$

Thus:

$$\forall c: \lim_{k \to \infty} W_c^{(k)} = [\![W_{\text{tot}}/C \pm 0.4\overline{9}]\!]. \quad \blacksquare \quad (10)$$

*Proposition 2 (Expected Delay until Convergence to SS):* For TFDMA with $W_{\text{tot}}$ nodes in $C$ channels, the expected delay (in s) until convergence to balanced state can be estimated by

$$d_{W_{\text{tot}},C} = T\left[\sum_{i=1}^{\frac{(W_{\text{tot}}+C-1)!}{(C-1)!W_{\text{tot}}!}} \left[p(i) \sum_{k=1}^{W_{\text{diff}}(i)} (d^{(k)} + 2)\right] + k_{\text{ss}}\right] \quad (11)$$

with: $i$ the index of the vector comprising a possible distribution of $W_{\text{tot}}$ nodes in $C$ channels (i.e. $[W_1(i) \dots W_C(i)]$,

$$p(i) = \prod_{c=1}^{C-1}\left[\binom{W_{\text{res},c}(i)}{W_c(i)} \frac{(C-1)^{W_{\text{res},c}(i) - W_c(i)}}{C^{W_{\text{res},c}(i)}}\right], \quad (12)$$

and $\quad d^{(k)} = \dfrac{1 - \left(1 - \beta^{k-1} p_{\text{sw},c}^{(0)}\right)^{Z(W_{\text{diff}}(i) + [\![W_{\text{tot}}/C]\!] - k + 1)}}{1 - \left(1 - \beta^{k-1} p_{\text{sw},c}^{(0)}\right)^{W_{\text{diff}}(i) + [\![W_{\text{tot}}/C]\!] - k + 1}}, \quad (13)$

with $\forall i: W_{\text{res},c}(i) = W_{\text{tot}} - \sum_{m=1}^{c-1} W_m(i)$, and

$$W_{\text{diff}}(i) = \max_{\forall c} |W_c(i) - [\![W_{\text{tot}}/C]\!]|. \quad (14)$$

*Proof:* When $W_{\text{tot}}$ nodes join $C$ channels randomly, the total number of combinations of nodes in channels, $L_{W_{\text{tot}},C}$, is:

$$L_{W_{\text{tot}},C} = (W_{\text{tot}} + C - 1)! / [(C-1)! W_{\text{tot}}!]. \quad (15)$$

The probability of each combination $i$ occurring is $p(i)$, given by (12), derived by iterating the binomial probability mass function for all channels $C$ (since nodes join channels randomly). Hence, the expected delay is

$$d_{W_{\text{tot}},C} = T \sum_{i=1}^{L_{W_{\text{tot}},C}} p(i) d_{\text{periods}}(i). \quad (16)$$

with $d_{\text{periods}}(i)$ the expected number of periods until convergence to SS is achieved for combination $i$. For each combination, $d_{\text{periods}}(i)$ is dominated by the channel with the largest imbalance from the average, since this channel will have the largest inflow or outflow of nodes. The largest imbalance is expressed by $W_{\text{diff}}(i)$ given by (14). The remainder of the proof estimates $d_{\text{periods}}(i)$.

We present the case of the channel with the largest surplus of nodes under combination $i$ (assumed to be $\text{Ch}\{c\}$); the equivalent hold for the channel with the largest deficit. Nodes will gradually leave $\text{Ch}\{c\}$ until $[\![W_{\text{tot}}/C \pm 0.4\overline{9}]\!]$ nodes remain in that channel. Since nodes decide independently on whether to attempt a switch, the probability that of no switching occurs within the first period is:

$$p_{\text{no\_sw},c}^{(0)} = \left(1 - p_{\text{sw},c}^{(0)}\right)^{W_{\text{diff}}(i) + [\![W_{\text{tot}}/C]\!]}, \quad (17)$$

By construction, one switching attempt must happen within (maximally) $Z$ periods. Hence, the expected number of periods until the first switching attempt takes place is:

$$d^{(1)} = \sum_{z=1}^{Z} z\left(p_{\text{no\_sw},c}^{(0)}\right)^{z-1}\left(1 - p_{\text{no\_sw},c}^{(0)}\right) + Z\left(p_{\text{no\_sw},c}^{(0)}\right)^Z \quad (18)$$

$$= \frac{1 - \left(p_{\text{no\_sw},c}^{(0)}\right)^Z}{1 - p_{\text{no\_sw},c}^{(0)}} \quad (19)$$

This is followed by two periods where nodes repeat their beacon message waiting for a "return" message. Iterating the above process, for the $k$th departure in $\text{Ch}\{c\}$, we reach $d^{(k)}$ given by (13). Finally, $d_{\text{periods}}(i)$ is found by the accumulation of all $W_{\text{diff}}(i)$ iterations, which leads to (11). $\blacksquare$

Proposition 2 demonstrates analytically the influence of design settings, $p_{\text{sw},c}^{(0)}$, $\beta$ and $k_{\text{ss}}$ (controlled by $\alpha$ [4]), as well as system parameters $C, W_{\text{tot}}, T$ and $Z$, on the expected delay.

## IV. EXPERIMENTS

### A. Experimental Setup

For our experiments, we used $W_{tot} = 16$ imote2 sensors (with the 2.4GHz Chipcon CC2420 wireless transceiver), placed in an obstacle-free topology. All messages used the TinyOS standard. The utilized parameters were: $q_{ss} = 0.02$, $T = 0.25s$, $\alpha = 0.95$, $\beta = 1.25$, $\forall c: p_{sw,c}^{(0)} = 0.33$, $s_c^{(0)} = 1$, $Z = 60$. Due to the use of higher convergence threshold than the one used in DESYNC, we found $k_{ss} = 6$, which leads to significantly-faster convergence to SS than what is reported in [4]. All measurements are averages of several trials of 60s each. Up to $C = 8$ channels were used (out of the 16 available in IEEE802.15.4), and one base station is used per channel to passively record all messages for subsequent analysis.

### B. Results and Comparisons

Table 1 contains the results with respect to bandwidth efficiency (the last column of the table is discussed separately in the next paragraph). We also present the results of DESYNC [4], TSMP [2] and the recently-proposed EM-MAC [8] in Table 2. These comprise the state-of-the-art in *centralized* [2] and *distributed* [8] channel hopping in WSNs. All approaches are realized over the same physical layer (IEEE802.15.4 and Chipcon CC2420 tranceiver). By comparing the two tables, it is evident that the total network throughput (throughput of all nodes) as well as the throughput *per node* is higher in the proposed TFDMA than in all other TDMA or channel hopping solutions when all 8 channels are used. Our throughput surpasses DESYNC even in the single channel case as we use higher convergence threshold, leading to faster convergence to SS. Unlike EM-MAC that is designed for low-bandwidth wireless transmissions *over lengthy periods of time*, the proposed TFDMA can achieve very high bandwidth for rapid message exchanges *within short intervals*. This is very suitable for WSN-based surveillance and monitoring, where infrequent alerts can initiate rapid wake-up and high volume of WSN traffic for short intervals, before the network suspends again.

| Total Channels | 1 | 2 | 4 | 8 | 8, hidden nodes & reshuffling |
|---|---|---|---|---|---|
| Tot. throughput (Kbps) | 126.9 | 266.7 | 543.8 | 801.9 | 649.0 |
| Max per node (Kbps) | 8.3 | 16.7 | 34.1 | 58.1 | 52.6 |
| Min per node (Kbps) | 7.3 | 16.5 | 33.7 | 43.5 | 32.1 |
| Message loss (%) | 0.54 | 0.01 | 0.01 | 0.96 | 0.98 |

Table 1. Throughput of the proposed TFDMA with 16 nodes.

| Protocol | DESYNC [4] | TSMP [2] | EM-MAC [8] |
|---|---|---|---|
| Tot. throughput (Kbps) | 55.0 | 574.4 | 5.1 |
| Max per node (Kbps) | 3.5 | 35.9 | 0.32 (average) |
| Min per node (Kbps) | 3.2 | (average) | 0.00 |
| Message loss (%) | 0.30 | 0.01 | 0.00 |

Table 2. Throughput obtained with DESYNC, TSMP and EM-MAC; all results are reported under a fully-connected WSN topology comprising 16 nodes.

We also measured the average time to achieve convergence to SS in TFDMA versus the estimate of Proposition 2 (Table 3). Table 4 shows the convergence time required by the other three solutions under comparison. Evidently, the proposed TFDMA achieves quick convergence, which agrees with the theoretical estimates of Proposition 2. Such low convergence times enable the application of node reshuffling (or suspension) in periodic intervals, i.e. all nodes can be forced to randomly join a new channel in order to increase their connectivity. By applying such node reshuffling every 60s, we obtained the results reported in the last column of Table 1; importantly, these results include the overhead of handling one-hop, possibly hidden, nodes based on the inclusion of neighboring nodes' beacon times within each node's beacon message, as proposed in [11]. These results still surpass the competing solutions despite the increase of beacon message size. A thorough study of properties of the proposed protocol under arbitrary topologies remains a topic for future work.

| Total Nodes | 16 | | | 8 | |
|---|---|---|---|---|---|
| Tot. Channels | 8 | 4 | 2 | 4 | 2 |
| Measured (s) | 4.7 [±1.7] | 4.0 [±1.0] | 3.2 [±0.5] | 3.1 [±0.7] | 2.9 [±0.6] |
| Proposition 2 (s) | 4.9 | 4.1 | 2.7 | 3.1 | 2.3 |

Table 3. Average delay (and standard error of mean) until SS.

| Protocol | DESYNC [4] | TSMP [2] | EM-MAC [8] |
|---|---|---|---|
| Delay until SS (s) | 8~48 | 48 | 8~9 |

Table 4. Average delay until SS under TSMP and EM-MAC.

## V. CONCLUSION

We proposed a new distributed time-frequency division multiple access protocol that utilizes the concept of reactive listening. Our approach distributes the available transmission opportunities in a balanced manner across time and frequencies (channels) in a sensor network without requiring the presence of a coordinator node. Stability and convergence time were derived analytically and then validated experimentally based on TinyOS imote2 wireless sensors. Our proposal allows for increased throughput and decreased convergence time versus TDMA-only schemes or versus centralized and distributed channel-hopping based approaches.